# Investigation of the frequency content of ground motions recorded during strong Vrancea earthquakes, based on deterministic and stochastic indices

Iolanda-Gabriela CRAIFALEANU[1, 2]

[1]Department of Reinforced Concrete Structures, Technical University of Civil Engineering Bucharest, Bd. Lacul Tei 124, sector 2, 020396 Bucharest, Romania

[2]National Institute for Research and Development in Construction, Urban Planning and Sustainable Spatial Development, URBAN-INCERC, INCERC Bucharest Branch, Sos. Pantelimon 266, sector 2, 021652 Bucharest, Romania

email: i.craifaleanu@gmail.com; iolanda@incerc2004.ro

ABSTRACT: The paper presents results from a recent study in progress, involving an extensive analysis, based on several deterministic and stochastic indices, of the frequency content of ground motions recorded during strong Vrancea seismic events. The study, continuing those initiated by Lungu et al. in the early nineties, aims to better reveal the characteristics of the analyzed ground motions. Over 300 accelerograms, recorded during the strong Vrancea seismic events mentioned above and recently re-digitized, are used in the study. Various analytical estimators of the frequency content, such as those based on Fourier spectra, power spectral density, response spectra and peak ground motion values are evaluated and compared. The results are correlated and validated by using the information provided by various spectral bandwidth measures, as the Vanmarcke and the Cartwright and Longuet-Higgins indices. The capacity of the analyzed estimators to describe the frequency content of the analyzed ground motions is assessed comparatively and the most relevant parameters are being identified.

## 1 INTRODUCTION

The seismicity of Romania is generated mainly by the Vrancea subcrustal source which has caused, over time, several destructive events. The first strong earthquake for which accelerographic records were obtained was that on March 4, 1977 (moment magnitude $M_w = 7.4$, focal depth $h = 94$ km), which caused losses of over 2 billion USD [1]. In the years that followed, three more earthquakes with $M_w > 6$ were generated by the same source. Due to the progressive extension of the seismic networks in Romania during this period, hundreds of accelerometric records were obtained for these strong earthquakes.

Several studies have shown important peculiarities of Vrancea events, e.g. significant differences in source mechanisms and directivity between events ([2], [3], [4] etc.) and a marked asymmetric distribution of the ground motion radiated by the Vrancea intermediate-depth earthquakes [5]. The causes of these peculiarities are not easy to discern. For instance, research aiming to separate site effects from other influences (source mechanism, propagation path) revealed that, for large areas located in earthquake-prone zones in Romania, bedrock is either difficult to identify or it is located at very large depths [6].

The paper presents results from a recent research in progress, involving a comprehensive analysis, based on the use of several deterministic and stochastic indices, of the frequency content of the ground motions recorded during strong Vrancea earthquakes. The research, continuing those initiated by Lungu et al. in [7] and developed afterwards in the late nineties [6], aims to better reveal the characteristics of the analyzed ground motions.

## 2 METHODOLOGY

### 2.1 Parameters considered

A relatively large number of parameters are used in the literature to characterize the frequency content of earthquake ground motions. Many of these are actually periods or frequencies related to the power spectral density (PSD) or to the response spectra. New and improved expressions of the scalar definitions of the frequency content, expressed as period values, were proposed in the last few years by different authors [8], [9], [10], [11]. In parallel, a number of bandwidth measures and related parameters are used as well [12].

The frequency content parameters used in the present study are described in the following.

#### 2.1.1 Parameters based on the power spectral density and on the Fourier amplitude spectrum

- The frequencies $f_1$, $f_2$ and $f_3$ corresponding to the first three peaks, in order of their amplitude, of the PSD [12] provide a set of straightforward indicators. Alternatively, the periods $T_1$, $T_2$ and $T_3$ can be used. In this paper, $T_1$ is used, under the notation of $T_{1(PSD)}$.
- Other simple indicators are the $f_{10}$, $f_{50}$ and $f_{90}$ fractile frequencies below which 10%, 50% and 90%, respectively, of the total cumulative power of PSD occur [12].
- The *central frequency*, $\Omega$, [13], is given by

$$\Omega = \sqrt{\frac{\lambda_2}{\lambda_0}} \quad (1)$$

The central frequency represents a measure of the frequency where the PSD is concentrated. In the above equation, $\lambda_2$ and $\lambda_0$ are 2nd and 0-th spectral moments of the one-sided spectral density of the stationary process of ground acceleration [12], given by the general expression

$$\lambda_n = \int_0^{\omega_n} \omega^n G(\omega) d\omega \qquad (2)$$

with $n = 2$ and 0, respectively, $\omega$ is the circular frequency and $G(\omega)$ is the spectral density function. Based on this definition, a central period, $T_{cen} = 2\pi/\Omega$, was computed for the present study.

- A related parameter is the *mean frequency* [14]

$$\overline{\omega} = \frac{\lambda_1}{\lambda_0} \qquad (3)$$

Based on this definition, a mean period, $T_{mean} = 2\pi/\Omega$, was computed for the present study.

- The *shape factor*, $q$ [13], is defined as

$$q = \sqrt{1 - \frac{\lambda_1^2}{\lambda_0 \lambda_2}}, \quad 0 \le q \le 1 \qquad (4)$$

The shape factor is being generally used as a measure of frequency bandwidth, as it reflects the dispersion of the power spectral density function about the central frequency. Higher values of this factor denote larger bandwidths.

- Another frequency bandwidth measure is the Cartwright and Longuet-Higgins $\varepsilon$ *parameter* [15], given by

$$\varepsilon = \sqrt{1 - \frac{\lambda_2^2}{\lambda_0 \lambda_4}}, \quad 0 \le \varepsilon \le 1 \qquad (5)$$

Wide frequency band processes are deemed to be those with $\varepsilon$ values close to 2/3 and smaller than 0.85. Narrow frequency band seismic processes have $\varepsilon$ values greater than 0.95 [12].

- A related parameter is $\xi$, proposed by D. M. Boore [16],

$$\xi = \sqrt{\frac{\lambda_2^2}{\lambda_0 \lambda_4}} = 1 - \varepsilon^2 \qquad (6)$$

- The *mean square period*, $T_{ms}$, proposed by E. M. Rathje [8], is defined as

$$T_{ms} = \left[ \sum_i C_i^2 \left( 1/f_i \right) \middle/ \sum_i C_i^2 \right] \text{ for } 0.25 \text{Hz} \le f_i \le 20 \text{Hz} \qquad (7)$$

where $C_i$ are the Fourier amplitudes of the accelerogram and $f_i$ are the discrete fast Fourier transform (FFT) frequencies between 0.25 to 20Hz. The frequency interval used in the discrete FFT computation, $\Delta f$, should be no greater than 0.05 Hz, to ensure that a stable value of $T_{ms}$ is obtained [9].

2.1.2 Parameters based on response spectra

- The *predominant period (based on acceleration spectrum)*, $T_{gSA}$, is defined as the period at which the maximum ordinate of an acceleration response spectrum calculated for 5% damping occurs [8].
- The *predominant period (based on velocity spectrum)*, $T_{gSV}$, is defined as the period at which the maximum ordinate of a velocity response spectrum calculated for 5% damping occurs [17].
- The *predominant period (based on input energy spectrum)*, $T_{gSEI}$, is defined as the period at which the maximum ordinate of an input energy response spectrum calculated for 5% damping occurs [18].
- The *characteristic period*, $T_1^*$, is defined as the period at the transition between the constant-acceleration and the constant-velocity segments of a 5% damped elastic spectrum, is given by

$$T_1^* = 2\pi \frac{(s_v)_{\max}}{(s_a)_{\max}} \qquad (8)$$

where $(s_v)_{\max}$ and $(s_a)_{\max}$ are the maximum ordinates in the pseudo-velocity and pseudo-acceleration response spectra, computed for 5% damping. An alternate definition of the characteristic period was proposed by Lungu et al. in 1997 [6], in the form

$$T_C = 2\pi\, EPV/EPA \qquad (9)$$

where

$$EPV = \left(SV_{\text{averaged on 0,4 s}}\right)_{\max}\!/2.5 \qquad (10)$$

$$EPA = \left(SA_{\text{averaged on 0,4 s}}\right)_{\max}\!/2.5 \qquad (11)$$

where $\left(SV_{\text{averaged on 0,4 s}}\right)_{\max}$ is the maximum value of the velocity response spectra averaged on a 0.4 s period mobile window and $\left(SA_{\text{averaged on 0,4 s}}\right)_{\max}$ is the maximum value of the acceleration response spectra averaged on a 0.4 s period mobile window.

The $T_C$ values used in this study were computed according to this second definition [21], [22].

- Ruiz-Garcia and Miranda [11] proposed modified definitions of the spectral moments $\lambda_n$, computed from the squared values of the velocity spectra, for elastic SDOF systems having damping ratio of 5%, as follows

$$\lambda_0^* = \sum_{i=1}^n S_{v,i}^2 \cdot \Delta T \qquad (12)$$

$$\lambda_1^* = \sum_{i=1}^n T_i \cdot S_{v,i}^2 \cdot \Delta T \qquad (13)$$

$$\lambda_2^* = \sum_{i=1}^n T_i^2 \cdot S_{v,i}^2 \cdot \Delta T \qquad (14)$$

Based on the above definitions, the cited authors computed a modified spectral characteristic period, $T_c^*$,

$$T_c^* = \frac{\lambda_1^*}{\lambda_0^*} = \frac{\sum_{i=1}^{n} T_i \cdot S_{v,i}^2 \cdot \Delta T}{\sum_{i=1}^{n} S_{v,i}^2 \cdot \Delta T} \quad (15)$$

and a modified central period

$$T_{cen}^* = \sqrt{\frac{\lambda_2^*}{\lambda_0^*}} = \sqrt{\frac{\sum_{i=1}^{n} T_i^2 \cdot S_{v,i}^2 \cdot \Delta T}{\sum_{i=1}^{n} S_{v,i}^2 \cdot \Delta T}} \quad (16)$$

where $n$ is the number of periods in the square velocity spectra, $S_v$ is the spectral velocity and $\Delta T$ is the period interval on the spectrum abscissa. Analogously to the Vanmarcke shape factor, $q$, they also computed a parameter, $\Omega$, defined as

$$\Omega = \sqrt{1 - \frac{(\lambda_1^*)^2}{\lambda_0^* \cdot \lambda_2^*}}, \quad 0 \leq \Omega \leq 1 \quad (17)$$

Like in the case of the original definition of the shape factor, smaller values of $\Omega$ are associated with narrow band signals (e.g. $\Omega < 0.5$ is considered a narrow band ground motion) [11].
To avoid any confusion with Vanmarcke's $\Omega$ parameter, the modified shape factor of Ruiz-Garcia and Miranda was denoted, in this paper, by $q^*$. For the same reason and also due to its analogy with $T_{mean}$, the modified spectral characteristic period, $T_c^*$, was be denoted by $T_{mean}^*$.

2.1.3  Parameters based on peak ground motion values

− An empirical relationship for the evaluation of the characteristic period of the ground motion was suggested by Heidebrecht in 1987 as

$$T_{4.3} = 4.3 \frac{PGV}{PGA} \quad (18)$$

and used by Fajfar and Fischinger in [19]. In the above equation, PGV is the peak ground velocity and PGA is the peak ground acceleration. This period was used to define the lower bound of the medium period region, i.e. the period range where the smoothed pseudo-velocity spectrum has its maximum values. According to the cited reference, the evaluation results from the assumption that spectral amplification in the acceleration-controlled region of the Newmark-Hall type spectrum is 1.46 times the spectral amplification in the velocity-controlled region.

## 2.2 Ground motion database

A database of 313 records from the four strong earthquakes (moment magnitude $M_w > 6$) generated by the Vrancea source during the last 34 years [23], including those from the destructive March 4, 1977 event, were considered in the study. The other three earthquakes were those on August 30, 1986 ($M_w$=7.1, focal depth $h$=131 km), May 30, 1990 ($M_w$=6.9, $h$=91 km) and May 31, 1990 ($M_w$=6.4, $h$=87 km).

The distribution of records with respect to each seismic event is shown in Fig. 1. The figure also displays the distribution for horizontal (H) and vertical (V) components.

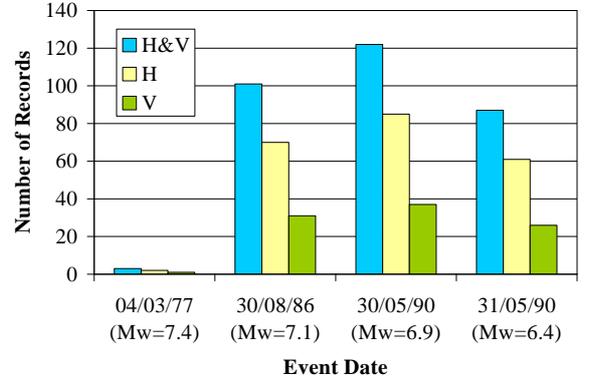

Figure 1. Distribution by events of the records in the database

The analysis was focused on the horizontal components of ground motions; however, the tendencies and correlations observed were checked as well for the total number of records and separately, for vertical records.

The records were also classified according to their frequency bandwidth, by using the criteria in Table 1, which are to some extent similar to those used by Lungu et al. in [6]. All the criteria in a row should be satisfied for a record to be classified as "narrow bandwidth" or "broad bandwidth".

Table 1. Criteria used in this study for the classification of ground motion records according to their bandwidth

| Bandwidth | $\varepsilon$ | $f_{50}$ | $T_C$ |
|---|---|---|---|
| Narrow | $\geq 0.95$ | $\leq 2.0$Hz | $\geq 0.95$s |
| Intermediate | all other records | | |
| Broad | $\leq 0.85$ | $\geq 3.0$Hz | $\leq 0.75$s |

By using the criteria in Table 1, of the 218 horizontal ground motion records considered in the study, 21 were identified as narrow band records, 151 as intermediate band records and 46 as broad band records.

## 3 CORRELATIONS BETWEEN THE GROUND MOTION FREQUENCY CONTENT PARAMETERS

Correlations between the seismic ground motion frequency content parameters were assessed. Of special interest were considered the correlations between the different period-type scalar parameters used for the characterization of the frequency content of seismic ground motions and also the correlations between these parameters and the bandwidth indicators, such as $q$, $q^*$, $\varepsilon$ and $\xi$.

Table 2 shows the correlation coefficients, $R^2$, computed for the horizontal components of the ground motions in the database. It was assumed that the relationship between two parameters is given by a linear functional, of the form $y = a \cdot x + b$.

Table 2. Correlation coefficients ($R^2$) between the analyzed ground motion frequency content and bandwidth parameters
(assumed relationship between two parameters: $y=a\cdot x+b$). Horizontal components for all seismic events

| | $T_{ms}$ | $T_{1(PSD)}$ | $T_{mean}$ | $T_{cen}$ | $T_{gSA}$ | $T_{gSV}$ | $T_{gSEI}$ | $T_C$ | $T_{mean}*$ | $T_{cen}*$ | $T_{4.3}$ | $q$ | $q*$ | $\varepsilon$ | $\xi$ |
|---|---|---|---|---|---|---|---|---|---|---|---|---|---|---|---|
| $T_{ms}$ | **1.000** | **0.531** | **0.763** | **0.570** | *0.441* | *0.285* | *0.417* | **0.831** | *0.330* | *0.235* | **0.759** | **0.515** | **0.574** | *0.445* | **0.551** |
| $T_{1(PSD)}$ | **0.531** | **1.000** | *0.384* | *0.261* | *0.227* | *0.189* | *0.358* | *0.468* | *0.242* | *0.182* | *0.378* | *0.324* | *0.375* | *0.161* | *0.201* |
| $T_{mean}$ | **0.763** | *0.384* | **1.000** | **0.921** | **0.543** | *0.127* | *0.204* | **0.512** | *0.111* | 0.066 | **0.559** | *0.272* | *0.267* | *0.481* | **0.588** |
| $T_{cen}$ | **0.570** | *0.261* | **0.921** | **1.000** | *0.443* | 0.071 | *0.126* | *0.341* | 0.048 | 0.023 | *0.410* | 0.098 | *0.149* | *0.420* | *0.491* |
| $T_{gSA}$ | *0.441* | *0.227* | **0.543** | *0.443* | **1.000** | 0.030 | 0.064 | *0.276* | 0.023 | 0.007 | *0.307* | *0.196* | *0.108* | *0.356* | *0.435* |
| $T_{gSV}$ | *0.285* | *0.189* | *0.127* | 0.071 | 0.030 | **1.000** | *0.417* | *0.366* | **0.627** | **0.574** | *0.266* | *0.253* | **0.565** | *0.127* | *0.133* |
| $T_{gSEI}$ | *0.417* | *0.358* | *0.204* | *0.126* | 0.064 | *0.417* | **1.000** | *0.426* | *0.403* | *0.338* | *0.318* | *0.258* | *0.474* | *0.111* | *0.127* |
| $T_C$ | **0.831** | *0.468* | **0.512** | *0.341* | *0.276* | *0.366* | *0.426* | **1.000** | *0.401* | *0.290* | **0.715** | *0.442* | **0.673** | *0.274* | *0.371* |
| $T_{mean}*$ | *0.330* | *0.242* | *0.111* | 0.048 | 0.023 | **0.627** | *0.403* | *0.401* | **1.000** | **0.975** | *0.356* | *0.229* | **0.769** | 0.053 | 0.072 |
| $T_{cen}*$ | *0.235* | *0.182* | 0.066 | 0.023 | 0.007 | **0.574** | *0.338* | *0.290* | **0.975** | **1.000** | *0.276* | *0.160* | **0.626** | 0.028 | 0.040 |
| $T_{4.3}$ | **0.759** | *0.378* | **0.559** | *0.410* | *0.307* | *0.266* | *0.318* | **0.715** | *0.356* | *0.276* | **1.000** | *0.379* | **0.519** | *0.310* | *0.372* |

As it can be observed from Table 2, in most cases (56%) correlation coefficients for the period-type parameters are in the intermediate range (0.1…0.5, values shown in italic font). In almost one third of the cases (31%) the correlation coefficients exceed 0.5 (values shown in bold font), while in about 13% of the analyzed cases a weak correlation is obtained, with $R^2$ values below 0.1 (values shown in normal font).

An analysis made on individual parameters shows that correlations differ substantially between the pairs of parameters taken into account. These differences originate basically from the parameter definitions. Some observations are briefly presented in the following.

The mean square period, $T_{ms}$, considered one of the most promising parameters for the characterization of ground motions frequency content [9], shows the best correlation ($R^2=0.831$) with $T_C$, the characteristic period based on modified definitions of effective peak ground motion values. Good correlations ($R^2 \cong 0.76$) are also observed with the mean period, $T_{mean}$ and with the characteristic period based on peak ground motion values, $T_{4.3}$. Lower correlation coefficients (0.570 and 0.531, respectively) result between $T_{ms}$ and $T_{cen}$ and between $T_{ms}$ and $T_{1(PSD)}$. The mean square period also shows a good correlation with the frequency bandwidth indicators, especially $q$, $q*$ and $\xi$.

The period corresponding to the maximum PSD value, $T_{1(PSD)}$, is rather poorly correlated with the other analyzed parameters, except $T_{ms}$.

The two periods based on the spectral moments of the PSD, $Tmean$ and $Tcen$, as well as their homologues $Tmean*$ and $Tcen*$ are well correlated to one another; however, between the two families of definitions, correlations are very poor. The alternate mean and central periods, $Tmean*$ and $Tcen*$, are also, predictably given their definition, well correlated to $T_{gSV}$. Both mentioned periods, and also as $T_{gSV}$, $T_C$ and $T_{4.3}$, are well-correlated with the modified shape factor, $q*$.

The predominant period based on the acceleration spectrum, $T_{gSA}$, shows generally moderate to poor correlation with the other analyzed parameters. The best correlation appears to be with $T_{mean}$ values ($R^2=0.543$).

Another parameter that is only moderately correlated with the other analyzed parameters is $T_{gSEI}$, for which the best correlation ($R^2=0.426$) occurs with $T_C$.

The $T_C$ and $T_{4.3}$ periods show good or intermediate correlation with all the other analyzed parameters. They are remarkably well correlated with the mean square period, $T_{ms}$.

The above correlations were also assessed separately for the 1986 and 1990 seismic events (the 1977 earthquake was not considered in this particular study, as only one complete record is available for this event). Figures 2 to 6 show some of the results obtained for the period-type parameters.

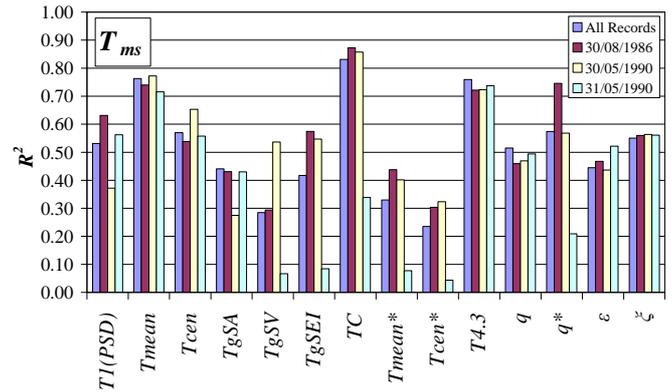

Figure 2. Correlation coefficients ($R^2$) between $T_{ms}$ and the other analyzed parameters

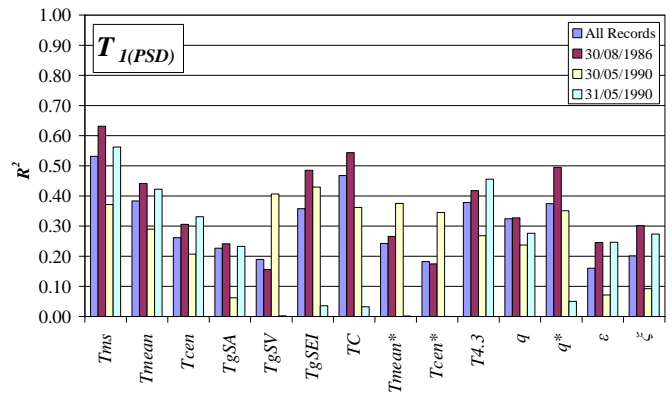

Figure 3. Correlation coefficients ($R^2$) between $T_{1(PSD)}$ and the other analyzed parameters

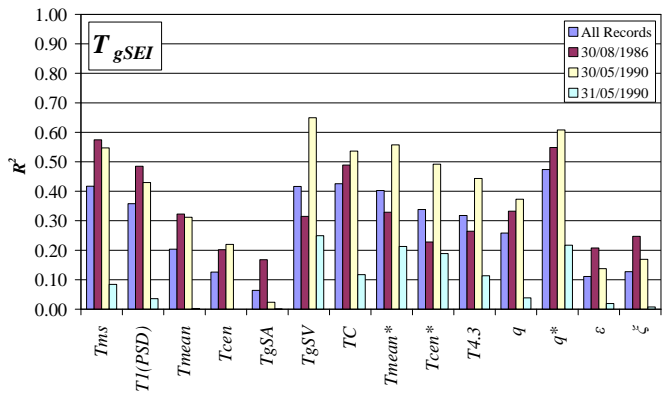

Figure 4. Correlation coefficients ($R^2$) between $T_{gSEI}$ and the other analyzed parameters

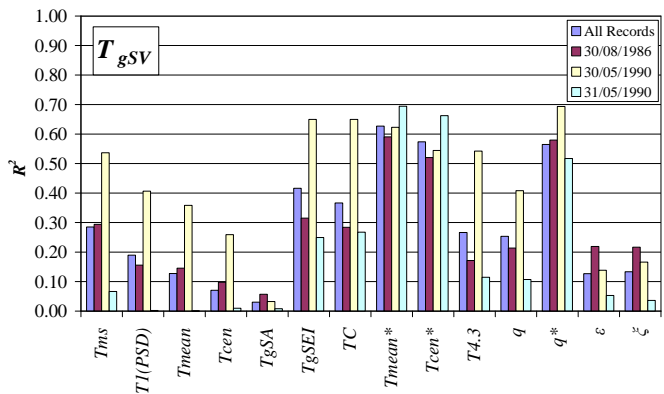

Figure 5. Correlation coefficients ($R^2$) between $T_{gSV}$ and the other analyzed parameters

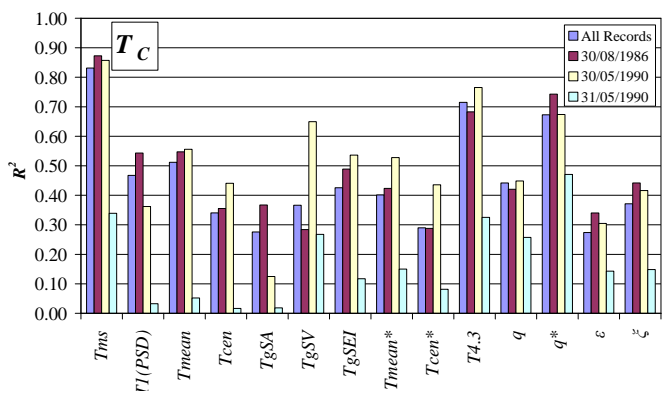

Figure 6. Correlation coefficients ($R^2$) between $T_C$ and the other analyzed parameters

It can be noticed, from the above figures, that the correlations differ significantly between the ground motion subsets corresponding to each seismic event.

For the 30/08/1986 and 30/05/1990 subsets of ground motions, the event-based separation of records leads, for most of the evaluations, to an improvement of the correlations between period-type parameters, by comparison to the evaluations made on the entire set of horizontal records. In the case of the 31/05/1990 subset, the same correlations tend to decrease in most of the evaluated cases.

It should be noted the good and relatively stable correlations, with respect to the ground motion set, between $T_{ms}$ and $T_{mean}$, $T_{cen}$, $T_c$, $T_{4.3}$ and $\xi$ (Fig. 2).

The type of functional chosen to express the relationship between the analyzed parameters can influence significantly the results. Fig. 7 shows the correlation between $T_{ms}$ and $T_C$, for all horizontal records and separately for the 1986 and 1990 seismic events, together with the corresponding regression equations. For this analysis, the functional was forced to pass through the origin, thus $y=a \cdot x$.

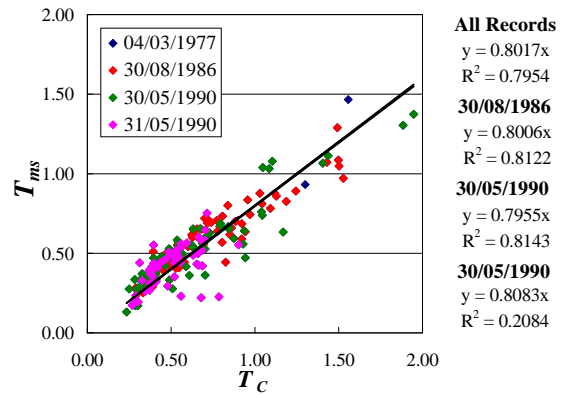

Figure 7. Correlation between $T_{ms}$ and $T_C$, for all horizontal records and separately for the 1986 and 1990 events (trendlines are forced to pass through the origin)

It can be noticed that the four trendlines are almost coincident, the regression coefficients being very close one to each other, as opposed to the differences shown in Fig. 2, where a functional of the form $y=a \cdot x+b$ was used.

The above result also suggests a stable correlation, with respect to ground motion subset, between $T_{ms}$ and $T_C$.

For the subsequent phases of the research, additional analyses are planned, including a study of vertical motions and the separate assessment of broad and narrow frequency bandwidth records.

A preliminary evaluation made on vertical records showed a better correlation between $T_{1(PSD)}$ and $T_{gSV}$, $T_{gSEI}$, $T_{mean}^*$ and $T_{cen}^*$, as compared with the set of horizontal records.

In what concerns the assessment of the analyzed ground motions based on their frequency bandwidth, due to the relatively small number of records available, only partial conclusions could be drawn on the correlation between the considered parameters. The good correlation observed previously between $T_{ms}$ and $T_C$, $T_{4.3}$ and $q^*$ was noticeable also for this set of motions. Improved correlations, as compared to the analysis of the entire set of horizontal ground motions, were observed between $T_{gSV}$ and $T_{gSEI}$, $T_{mean}^*$ and $T_{cen}^*$. This appears to sustain the suggestion of Ruiz-Garcia and Miranda [11] concerning the possible use of velocity spectrum-related parameters in describing the frequency content of narrow frequency band motions.

4 CONCLUSIONS

The capacity of eleven different scalar period-type parameters, used in the literature, to describe the frequency content of ground motions was assessed comparatively, by using a database of over 300 records obtained from four strong Vrancea earthquakes with moment magnitude larger than 6. Additionally, information provided by a number of spectral bandwidth measures was used.

Correlations between the considered parameters were determined for the entire ground motion set, as well as for subsets created by type of component (horizontal or vertical), event or frequency bandwidth.

One of the most reliable scalar parameter appeared to be the mean square period, $T_{ms}$, proposed by Rathje et al. in 1998 [8]. This parameter, defined as a function of Fourier amplitudes and frequencies, showed good and relatively stable correlations with other period-type parameters, based either on power spectral density, response spectra or peak ground motion values, as well as with the frequency bandwidth measures. Other promising period-type scalar parameters were found to be the characteristic period based on modified definitions of the effective peak ground motion values, proposed by Lungu et al. in 1997 [6], the predominant period proposed by Heidebrecht in 1987 [19] and, possibly, the mean and the central period, based of the spectral moments of power spectral density.

Further studies will be needed to clarify, in the subsequent phases of the research, the dependency of the frequency content of the analyzed motions on different factors, such as the directivity of seismic wave propagation or site conditions.

ACKNOWLEDGMENTS

The author would like to thank Dr. Ioan Sorin Borcia from URBAN-INCERC, INCERC Bucharest Branch, for providing the $T_C$ values for the analyzed ground motion records.
The work reported in this paper was partly sponsored by the Romanian National Authority for Scientific Research, in the framework of the "Nucleu" Programme, Project No. PN 09-14.01.03.